\documentclass[pdflatex,sn-nature]{sn-jnl}
\usepackage[T1]{fontenc}
\usepackage{lmodern}
\usepackage{graphicx}
\usepackage{amsmath,amssymb}
\usepackage{xcolor}
\usepackage{booktabs}
\usepackage{url}
\usepackage{floatpag}

\begin{document}

\title{Accelerating wave-scattering simulations with learned search directions}

\author[1]{\fnm{Haoran} \sur{Sun}}
\author[2]{\fnm{Tao} \sur{Shan}}
\author[3]{\fnm{Hongyu} \sur{Zhou}}
\author[4]{\fnm{Rui} \sur{Guo}}
\author[1]{\fnm{Shenheng} \sur{Xu}}
\author[1]{\fnm{Fan} \sur{Yang}}
\author*[1]{\fnm{Maokun} \sur{Li}}\email{maokunli@tsinghua.edu.cn}
\affil[1]{\orgdiv{Department of Electronic Engineering}, \orgname{Tsinghua University}, \orgaddress{\city{Beijing}, \country{China}}}
\affil[2]{\orgdiv{School of Electronic and Information Engineering}, \orgname{Beihang University}, \orgaddress{\city{Beijing}, \country{China}}}
\affil[3]{\orgname{Institut Langevin, ESPCI Paris, PSL University}, \orgaddress{\city{Paris}, \country{France}}}
\affil[4]{\orgdiv{Faculty of Mathematics and Computer Science}, \orgname{Weizmann Institute of Science}, \orgaddress{\city{Rehovot}, \country{Israel}}}

\abstract{Wave-scattering simulation requires solving a dense boundary-element system for every frequency, incidence and design. Krylov solvers build each search space from scratch, and it grows with problem size. Here we present BEMFormer, a solver that learns to generate search directions across systems of one integral equation and supplies geometry-specific directions through a coarse-grid correction. BEMFormer generalizes to unseen geometries, meshes and excitations, and to acoustic scattering after fine-tuning on eight geometries. It reproduces the measured radar cross-sections of benchmark targets. On a quadcopter with 1.04 million unknowns, 38 times the largest training problem, BEMFormer solves each system 30 times faster than a preconditioned Krylov solver. Both reach a radar cross-section deviation of about 1~dB, comparable to the difference between independent measurements and acceptable for engineering applications. These results show that solving million-unknown scattering problems in minutes is possible when the search space is learned rather than built from scratch.}

\maketitle

Numerical simulation of wave scattering is essential for radar and sonar\cite{Harrington68,Wilkes20}, wireless communication and the design of microwave and acoustic devices, including metasurfaces\cite{CuiCoding14}. For a scatterer in free space, the problem can be formulated as a boundary integral equation on its surface, and the boundary element method discretizes this equation into a dense, complex-valued and non-Hermitian linear system\cite{Gibson08}. A design study requires one solution for every frequency, incidence direction and shape, and data-driven design methods require many such solutions for training and validation\cite{LiMetaAI26,HouInfoMetaGen26}. Engineering targets are often many wavelengths in size, and their discretizations have hundreds of thousands to millions of unknowns. Solving many large, dense systems efficiently therefore remains a central challenge in scattering simulation.

Classical methods solve such systems either directly or iteratively. Direct methods compute a lower--upper (LU) factorization, which can be reused for every incidence direction\cite{Shaeffer08,Jiang22}. Its cost, however, is cubic and its memory usage is quadratic in the number of unknowns. Larger problems are solved iteratively, for example by the generalized minimal residual method (GMRES), which minimizes the residual over a Krylov subspace that grows by one dimension per iteration\cite{SaadSchultz86}. Fast algorithms such as the multilevel fast multipole algorithm (MLFMA) compute each matrix--vector multiplication at a fraction of the cost of a dense multiplication\cite{CRW93,SLC97}. With the MLFMA, problems with millions or even billions of unknowns have been solved\cite{PanBillion12,Michiels15}. The number of iterations, however, grows with the electrical size of the object, its size measured in wavelengths\cite{Pocock97}. Preconditioners, deflation and Krylov subspace recycling reduce this number\cite{MalasGurel07,Andriulli08,CDG03,Parks06}, but each is built from the matrix at hand and can be reused only for the same or closely related matrices. Every new geometry or frequency thus starts from scratch.

Problems governed by the same integral equation, however, form a family and share the same physics at the wavelength scale. A solver that has learned from many of them could therefore use a much smaller search space. Surrogate models and neural operators predict the solution directly from the geometry and excitation\cite{FNO21,Transolver24,DIMON24,PhiGRL24,PdEgatSCL25,KongEMPN24,ZhouJiao24}. On unseen geometries, their errors typically remain well above what engineering applications tolerate, with no means of reducing them. Hybrid methods keep a classical solver and learn one of its components, such as the preconditioner\cite{HINTS24,GNP25,Azulay22}, the search directions\cite{DCDM23,GRANS26,RINS26} or the recycled subspace\cite{NeurKItt24}. Because the exact operator remains in the loop, their accuracy improves with each iteration. So far, these methods have been applied to sparse linear systems. No learned solver has yet been trained across geometries for the dense linear systems of boundary integral equations and shown to be faster than preconditioned fast multipole solvers at the same accuracy.

In this work, we introduce BEMFormer, a boundary-element solver that learns its search directions from a family of scattering problems. It has two components. The first is a transformer that, in each iteration block, takes the residual, the geometry and the incident field as input and generates three search directions, in place of a Krylov subspace. The exact operator is applied to these directions and their coefficients are obtained by least squares, so the residual is computed rather than predicted. However, error along the few near-null directions of each geometry produces almost no residual, so the network receives almost no learning signal to remove it. The second component, a coarse-grid correction, supplies these geometry-specific directions at low cost.

We show that BEMFormer generalizes to unseen geometries, meshes and excitations, and that it can be transferred to acoustic scattering by fine-tuning on only eight geometries. Its electromagnetic solutions reproduce the measured radar cross-sections (RCS) of the NASA almond and other benchmark targets. Trained only on problems with up to about 27,000 unknowns, BEMFormer solves a quadcopter problem with 1.04 million unknowns 30 times faster than GMRES with a block-Jacobi preconditioner at the same RCS accuracy.

\section*{Results}

\subsection*{Overview of BEMFormer}\label{sec:r1}
A family of scattering problems comprises the linear systems $Ax=b$ that arise from one integral equation for different closed surfaces, frequencies and incident fields (Fig.~\ref{fig:1}a). Most of the singular-value spectrum is common to the systems of a family, whereas the few smallest singular values depend on the specific geometry (Fig.~\ref{fig:1}b). They arise from cavity resonances and from nearly touching parts, and we call the corresponding singular directions the near-null directions. BEMFormer therefore combines two components (Fig.~\ref{fig:1}c). A network learns from the family to generate search directions in 25 iteration blocks, and a coarse-grid correction supplies the near-null directions of each geometry.

A Krylov method seeks the solution in the affine space
\begin{equation}
x_m\in x_0+\mathcal K_m(A,r_0),\qquad \mathcal K_m(A,r_0)=\mathrm{span}\{r_0,Ar_0,\dots,A^{m-1}r_0\},
\label{eq:krylov}
\end{equation}
where $r_0=b-Ax_0$ and $\mathcal K_m$ is the Krylov subspace. GMRES selects the point in this space with the smallest residual\cite{SaadSchultz86}. BEMFormer replaces this space with the span of directions that a network generates (Fig.~\ref{fig:2}a). Block $t$ updates the solution as
\begin{equation}
\begin{aligned}
x_{t+1}&=x_t+\mathcal P_tc_t,\qquad c_t=\arg\min_{c}\,\|r_t-A\mathcal P_tc\|,\qquad \mathcal P_t=[P_0,\dots,P_t],\\
P_t&=f_{\theta_t}\bigl(r_t,\,Ar_t,\,A\Delta x_{t-1},\,A\Delta x_{t-2},\,x_t,\,b,\,\hat k;\,\Gamma\bigr),
\end{aligned}
\label{eq:block}
\end{equation}
where $r_t=b-Ax_t$, $P_t$ contains the three directions generated by the network $f_{\theta_t}$ of block $t$, $\Delta x_{t-1}$ and $\Delta x_{t-2}$ are the two previous updates, $\hat k$ is the incidence direction and $\Gamma$ is the surface mesh. Each block has its own weights. In equation~(\ref{eq:krylov}), the search space is generated by powers of $A$ acting on $r_0$, and in equation~(\ref{eq:block}) by a learned map that also takes the geometry and the incident field as input. The network thus generates only the search space, whereas the residual is always evaluated with the exact operator. From $x_0=b$, one solution requires 101 matrix--vector multiplications.

The network is a transformer whose tokens are the basis functions of the mesh (Fig.~\ref{fig:2}b). A phase-rotation encoding, a variant of rotary position encoding\cite{RoFormer24} that uses the phases of plane waves, makes attention depend on the phase difference between two basis functions. Each block has two encoder layers, whose attention has three branches that mirror the near-field and far-field interactions of a fast multipole method (Fig.~\ref{fig:2}c). The basis functions are grouped into clusters. Local-window attention connects each cluster with its neighbouring basis functions and acts as the near-field interactions. A summary branch attends to a single vector for each cluster, and a selection branch attends to all basis functions of the clusters with the highest attention scores in the summary branch. Together, the summary and selection branches represent the far-field interactions, at low resolution for most clusters and at full resolution for the selected clusters. The order in which the unknowns are numbered is arbitrary, and the network is permutation equivariant by construction. Both the phase-rotation encoding and the clusters are computed from the positions of the basis functions, and a random reordering leaves the solution unchanged (Supplementary Fig.~1a and Supplementary Note~1).

For the coarse-grid correction, a coarse mesh of the surface is generated, and inverse subspace iteration on the coarse operator yields up to 64 near-null vectors (Fig.~\ref{fig:2}d). They are interpolated to the fine mesh and are used in every block through a two-sided projection, a form of deflation\cite{CDG03} built into the least-squares step. This computation is performed once per geometry and shared by all incidences (Methods). The network removes most of the error, and the correction removes the error along the near-null directions (Fig.~\ref{fig:2}e).

BEMFormer was trained on 2,408 geometries with up to 27,180 unknowns, at one frequency and with one polarization (Methods, Supplementary Fig.~2 and Supplementary Note~2). We tested BEMFormer on targets outside the training distribution in size, shape and excitation, up to 1.04 million unknowns (Fig.~\ref{fig:1}d).

\subsection*{Accuracy on unseen geometries}\label{sec:r3}
BEMFormer is designed to solve scattering problems for geometries and excitations not seen during training, with a fixed number of matrix--vector multiplications. To assess its accuracy and cost, we compared it with three classical solvers and three learned surrogate models (Methods). The classical solvers are GMRES with a diagonal, a block-Jacobi or a near-field sparse-LU preconditioner, and we refer to them as GMRES, block-Jacobi GMRES and near-field-LU GMRES. In all tests, we quantify the cost of a solution by the number of matrix--vector multiplications with $A$ and measure wall-clock time separately. Accuracy is measured by the relative error against a reference solution and by the mean RCS deviation. Two independent measurements of the NASA almond\cite{Woo93,Kelley20} differ by 0.62--1.36~dB under this metric. We therefore consider an RCS deviation below 1~dB acceptable for engineering applications (Methods).

We first assessed accuracy on new shapes from the training distribution, the 687 geometries of the test set (Methods). With 101 matrix--vector multiplications, the median relative error is 0.024\% for the network alone and 0.018\% for BEMFormer (Fig.~\ref{fig:3}a). The relative error is below 1\% in 99.4\% of the systems for BEMFormer, compared with 64.8\% for block-Jacobi GMRES with the same number of multiplications. After a single forward pass, the median relative errors of the three surrogate models, Transolver\cite{Transolver24}, PhiGRL\cite{PhiGRL24} and PdEgatSCL\cite{PdEgatSCL25}, exceed 20\%.

We then examined the contribution of each component. We assessed the coarse-grid correction on ill-conditioned problems, for which the near-null directions are most important. On the test geometries with a condition number above 50, the median relative error of the network alone is 1.1\%, and the correction reduces it to 0.041\% (Fig.~\ref{fig:3}a). In this group, the eight smallest singular directions account for most of the error of the network alone, and with the correction this fraction is below one-fifth (Fig.~\ref{fig:2}e). For the network, ablation experiments that removed one component at a time showed that the least-squares update and the phase-rotation encoding are essential (Extended Data Fig.~\ref{ed:2} and Supplementary Note~3). The selection branch improves accuracy only on problems with 270,000 unknowns or more.

The learned search space is more effective than the Krylov subspace (Fig.~\ref{fig:3}b). With a median of only 12 directions, it reaches the residual that GMRES attains after 75 iterations. The number of matrix--vector multiplications that BEMFormer requires to reach a 1\% residual also remains nearly constant across the test set, whereas that of block-Jacobi GMRES increases with the number of unknowns. On a quadcopter larger than any training geometry, block-Jacobi GMRES already requires five times as many multiplications as BEMFormer to reach a 3\% error, with or without the same coarse space (Fig.~\ref{fig:3}c). On problems of up to 22,000 unknowns, however, classical solvers are competitive with BEMFormer and faster in wall-clock time (Extended Data Fig.~\ref{ed:3} and Supplementary Note~4). BEMFormer becomes faster than the classical solvers on larger problems, as shown below. These results show that, for unseen geometries from the training distribution, a fixed number of matrix--vector multiplications is sufficient for an accurate solution.

\subsection*{Generalization to unseen meshes, frequencies and excitations}\label{sec:r3b}
In practice, the mesh, the frequency and the excitation often differ from those of the training problems. We tested BEMFormer on up to 200 test geometries with other meshes and frequencies, with horizontal polarization and with an electric dipole or a line current as the source (Fig.~\ref{fig:3}d and Extended Data Fig.~\ref{ed:5}). The meshes have 10 to 25 elements per wavelength, the frequencies are 0.5 and 2~GHz, and the sources are placed one wavelength from the target. The median relative error of the network alone is 0.019--0.32\%, compared with 0.025\% for the training settings. With the correction, the median relative error is 0.015--0.17\%, and the RCS deviation is below 1~dB for all systems of these sets (Supplementary Note~5).

Radar-signature analysis requires the RCS of complex targets, which are often larger than the training geometries. We applied BEMFormer to three representative targets, the quadcopter, a satellite and a fixed-wing drone (Fig.~\ref{fig:3}e). For the quadcopter, we computed a full monostatic sweep in both polarizations. For the satellite and the drone, the MLFMA operator was used, with eight incidences each. For all three targets, the mean RCS deviation of BEMFormer from the reference is below 0.5~dB, whereas block-Jacobi GMRES does not converge with the same number of multiplications.

Finally, we changed the combination coefficient $\alpha$ of the integral equation to test whether the learned directions depend on the equation (Methods). Accuracy decreases when the equation changes (Fig.~\ref{fig:3}d). The median relative error of the network alone increases to 1.9\% for $\alpha=0.3$ and to 16\% for the magnetic-field integral equation alone ($\alpha=0$). For the electric-field integral equation alone ($\alpha=1$), it is 18.0\% (Extended Data Fig.~\ref{ed:5}c). This indicates that the network has learned to solve the systems of one integral equation, so a different equation requires fine-tuning. After fine-tuning on the electric-field integral equation, the median relative error of the network alone decreases to 3.97\%. Together, these results show that BEMFormer generalizes across geometries, meshes and excitations within one integral equation.

\subsection*{Transfer to acoustic scattering}\label{sec:r4}
The integral equations of acoustic and electromagnetic scattering share the same free-space Green's function. To test whether the learned search directions can be transferred between the two problems, we replaced the combined-field integral equation by its acoustic counterpart, the Burton--Miller equation for scattering by rigid, or sound-hard, surfaces\cite{BM71}. The unknowns are the pressures at the vertices instead of the currents on the edges (Fig.~\ref{fig:4}a). The network architecture and the least-squares update remain unchanged, and fine-tuning starts from the electromagnetic weights. The test set has 200 held-out geometries at 1,144~Hz. We fine-tuned the network separately on 8, 32, 128 and 512 geometries from the electromagnetic training set (Methods).

Without fine-tuning, the median relative error of the network is 37.5\% (Fig.~\ref{fig:4}b). Fine-tuning on only eight geometries reduces it to 1.3\%, and the coarse-grid correction reduces it further to 0.0012\%. For the same geometries and number of optimizer steps, fine-tuning is more accurate than training the same architecture from scratch (Supplementary Note~6). The error of the network alone decreases with the number of fine-tuning geometries (Fig.~\ref{fig:4}b). To balance the accuracy and the number of geometries, we used the network fine-tuned on 32 geometries in the following tests. Its median relative error is 0.91\% for the network alone and 0.0018\% for BEMFormer (Fig.~\ref{fig:4}c).

We then compared BEMFormer with GMRES on the test set and on more challenging problems (Fig.~\ref{fig:4}c). On the test set, the median relative error of both methods is below 0.002\%. The relative error is below 1\% in all systems for BEMFormer, compared with 68.3\% of the systems for GMRES. At twice the frequency, the median relative error is 0.014\% for BEMFormer and 13\% for GMRES. We also tested five complex geometries that are larger than the largest fine-tuning problem (Fig.~\ref{fig:4}d). Their median relative error is 6.0\% for the network alone and 0.20\% for BEMFormer, compared with 32.5\% for GMRES. Without fine-tuning, the same correction leaves a median relative error of 32\% on these geometries, so the accuracy stems from the fine-tuned network (Supplementary Note~6).

For comparison with measurements, we computed head-related transfer functions, the sound pressures at both ears as functions of the source direction, for the FABIAN head-and-torso simulator\cite{Brinkmann17,FABIANdb} (Fig.~\ref{fig:4}e). This problem has 14 times as many unknowns as the largest fine-tuning problem. In the horizontal plane, the computed transfer functions deviate from the measurement by a median of 0.69~dB, compared with 0.58~dB for LU factorization of the dense matrix on the same mesh (dense LU). The deviation can therefore be attributed to the mesh resolution rather than to the iterative solver (Supplementary Note~6). These results show that the learned search directions are not specific to electromagnetics and can be transferred to acoustic scattering with a few geometries.

\subsection*{Validation against measurements}\label{sec:r5}
In the previous sections, BEMFormer was compared with reference solutions on the same mesh. To test the complete simulation, including the discretization, we compared the computed monostatic RCS with measurements of the benchmark targets of the Electromagnetic Code Consortium (EMCC)\cite{Woo93}, such as the NASA almond (Fig.~\ref{fig:5}a and Supplementary Fig.~3). The meshes have 12.5 elements per wavelength. For comparison, we used two independent measurements of the almond\cite{Kelley20} and the results of another full-wave solver, ARCHIE-AIM\cite{Massey18}.

For the almond in vertical polarization, the simulation deviates from the two measurements by 0.72--1.32~dB, comparable to the difference of 0.99--1.12~dB between the two measurements (Fig.~\ref{fig:5}b,c). The computed RCS is also in good agreement with that of ARCHIE-AIM, with a difference not exceeding 0.6~dB. The iterative solver has little effect on these deviations. In the cases small enough for dense LU, the mean RCS deviation between BEMFormer and dense LU does not exceed 0.021~dB.

The deviations are larger in horizontal polarization and for the other targets, whose measured curves were extracted from published figures (Supplementary Note~7). They can be attributed to the discretization and to measurement uncertainty (Fig.~\ref{fig:5}d). When the mesh density is doubled, the computed RCS is closer to the measurement in 25 of 26 cases, and the median deviation decreases from 2.2 to 1.8~dB (Fig.~\ref{fig:5}b).

A comparison with an analytical solution removes measurement error altogether (Fig.~\ref{fig:5}e). For a perfectly conducting sphere, the mean deviation from the Mie series\cite{BohrenHuffman83} decreases from at most 0.36~dB to at most 0.12~dB when the mesh density is doubled. BEMFormer differs from dense LU by less than 0.01~dB, even at the five interior resonances of the sphere. These results show that the accuracy of the complete simulation is limited by the discretization and by measurement uncertainty, not by BEMFormer.

\subsection*{Scaling to a quadcopter with one million unknowns}\label{sec:r6}
BEMFormer is trained on small problems, whereas engineering targets are much larger. To assess how its cost and accuracy change with problem size, we simulated the same quadcopter at higher frequencies, with 41,000 to 1.04 million unknowns (Fig.~\ref{fig:6}a). The network was applied without retraining or fine-tuning. Each classical solver was run until it reached the RCS deviation of BEMFormer (Methods).

As the problem size increases, the dimension of the BEMFormer search space remains constant, whereas the Krylov subspace grows (Fig.~\ref{fig:6}b). BEMFormer uses 139 search directions at all sizes, whereas block-Jacobi GMRES requires 325 to 8,720. As a result, the solution time of BEMFormer increases more slowly with the number of unknowns than that of the classical solvers. The end-to-end times, including matrix assembly and the one-time preparation of each method, are shown in Fig.~\ref{fig:6}c,d. For small problems, however, dense LU or near-field-LU GMRES can be faster when many incidences are required (Fig.~\ref{fig:6}e).

The accuracy decreases as the electrical size increases, and the median relative error increases from 0.53\% for the smallest problem to 19.8\% for the largest (Fig.~\ref{fig:6}f). Relative to a reference solution from converged block-Jacobi GMRES (Methods), the median RCS deviation remains below 1~dB up to 736,000 unknowns. The fixed-wing drone of Fig.~\ref{fig:3}e follows a similar trend up to 1.03 million unknowns (Supplementary Note~8). Increasing the model size improves the accuracy (Supplementary Note~8).

The search directions of BEMFormer are generated by a network trained on a family of systems, and BEMFormer requires no additional iterations as the problem size increases. Two observations explain this (Supplementary Note~9). The fraction of the solution that each block contributes is similar at all sizes. The response of a block to a perturbation of the residual at one point is local, and more than 96\% of its energy typically lies within half a wavelength of that point.

The largest problem is the most challenging in this study and the closest to an engineering simulation. It has 1.04 million unknowns and an electrical size of 41 wavelengths. By comparison, the training problems do not exceed 27,180 unknowns and 5.5 wavelengths. The largest problem is therefore 38 times larger than the largest training problem in the number of unknowns and 7.4 times larger in electrical size. For this problem, BEMFormer is 30.5 times faster than block-Jacobi GMRES, the fastest classical solver (Fig.~\ref{fig:6}b). For the 361 incidences of a monostatic RCS curve, the end-to-end time is 16~h for BEMFormer and 21~days for block-Jacobi GMRES. The median RCS deviation of BEMFormer is 1.24~dB, comparable to the difference between two independent measurements of the almond (Fig.~\ref{fig:6}g). The deviation is most pronounced at the observation angles of weak scattering. Over the quarter of the angles with the strongest scattering, it is 0.57~dB (Supplementary Note~8). These results show that a search space learned on small problems remains effective on problems much larger than those seen during training.

\section*{Discussion}
We have introduced BEMFormer, a boundary-element solver that learns from a family of scattering problems how to generate its search directions. As a result, the search space of a dense scattering system is no longer built from scratch, and the cost of training is shared across all systems of the family. BEMFormer generalizes to unseen geometries, meshes, frequencies and excitations with a fixed number of matrix--vector multiplications. It can be transferred to acoustic scattering by fine-tuning on only eight geometries. Its solutions reproduce the measured RCS of benchmark targets, with an accuracy limited by the discretization and by measurement uncertainty rather than by the solver. On a quadcopter with 1.04 million unknowns, the search space of BEMFormer has 139 directions, whereas block-Jacobi GMRES requires 8,720 directions to reach the same RCS deviation. BEMFormer therefore solves each system 30 times faster than block-Jacobi GMRES.

The learned component provides directions outside the Krylov subspace. A Krylov method can only apply polynomials in $A$ to the initial residual, whereas the network computes a nonlinear function, learned from the family, of the residual, the geometry and the incident field. The correction component supplies, at low cost, the near-null directions, which training on the residual does not capture. Unlike surrogate models, which predict the solution in a single step, BEMFormer evaluates the residual of every solution with the exact operator, so its accuracy can be verified. Unlike learned preconditioners, which still rely on a Krylov iteration, BEMFormer generates the search directions themselves, so every solution requires the same number of matrix--vector multiplications.

The response of each block to the residual is local (Supplementary Note~9), and the coupling between distant parts of the surface comes from the exact operator. A local map does not depend on the size of the object. This is consistent with the generalization from 27,000 to 1.04 million unknowns. The transfer to acoustic scattering also shows that the learned weights are a starting point for a different integral equation.

BEMFormer requires only products with the operator, so it is compatible with existing fast algorithms such as the MLFMA\cite{SLC97} and adaptive cross approximation\cite{ZhaoACA05}. It can serve as a fast forward solver for electrically large problems and as a simulation engine for tasks that require many solutions, such as inverse design and shape optimization. BEMFormer is well suited to problems with about 77,000 unknowns or more, which corresponds to about 11 wavelengths on the quadcopter. For these problems, it was the fastest method in our experiments. Because the network is trained only once for a whole family, its training cost is recovered in studies that require many solutions (Supplementary Note~8).

BEMFormer has several limitations. First, the accuracy is lower for problems far outside the training distribution. A larger model is more accurate, but the gain is smaller for problems far from the training sizes (Supplementary Note~8). Second, the training set has a single frequency, polarization and set of incidence directions. In the monostatic sweep of the quadcopter, the median relative error for horizontal polarization is about five times higher than for vertical polarization (Supplementary Note~5). Training with several frequencies and polarizations may improve both accuracy and generalization. Finally, BEMFormer solves the discretized equations, so it does not reduce the discretization error (Fig.~\ref{fig:5}d).

The approach could be extended to dielectric bodies, radiation problems and other wave equations, such as those of elastodynamics, and to larger problems and networks. More broadly, our results suggest that a network trained on a family of linear systems can generate a search space that is more effective than a Krylov subspace, achieving a given accuracy with fewer directions.

\section*{Methods}

\subsection*{Boundary-integral formulation and discretization}
We use the boundary element method, known in electromagnetics as the method of moments\cite{Harrington68,Gibson08}.

\paragraph{Electromagnetic scattering.}
For a perfectly conducting closed surface $\Gamma$ with outward normal $\hat n$, the surface current $\mathbf J$ satisfies the combined-field integral equation
\begin{equation}
\mathcal L\mathbf J=\alpha\,e^{\mathrm i}-(1-\alpha)\,\eta\,h^{\mathrm i},\qquad \mathcal L=\alpha\,\mathcal L_{\mathrm E}-(1-\alpha)\,\eta\,\mathcal L_{\mathrm H},\qquad \alpha=0.5,
\label{eq:cfie}
\end{equation}
where $\mathcal L_{\mathrm E}$ and $\mathcal L_{\mathrm H}$ are the electric- and magnetic-field integral operators with the free-space Green's function and $\eta$ is the wave impedance. The right-hand sides $e^{\mathrm i}$ and $h^{\mathrm i}$ of the electric- and magnetic-field integral equations are formed from the incident fields. A time dependence of $e^{\mathrm i\omega t}$ is assumed and suppressed. The free-space Green's function is $e^{-\mathrm ikR}/(4\pi R)$, and the incident plane wave arrives from the direction $(\theta,\varphi)$. The current is expanded in Rao--Wilton--Glisson basis functions\cite{RWG82}, $\mathbf J=\sum_nx_n\mathbf f_n$, and Galerkin testing yields $A_{mn}=\langle\mathbf f_m,\mathcal L\mathbf f_n\rangle$, with $\mathcal L$ from equation~(\ref{eq:cfie}). Integrals over the source triangles are computed with nine quadrature points, and the testing integrals with one point at the centroid. Pairs of triangles closer than one tenth of a wavelength are integrated with the Duffy transformation, and the singular part of the magnetic-field kernel is subtracted and integrated analytically. The matrix is assembled in complex double precision on a graphics processing unit (GPU) and stored in single precision. All surfaces are meshed uniformly with Gmsh\cite{Gmsh09}. The mesh density in elements per wavelength is the wavelength divided by the mean edge length.

\paragraph{Acoustic scattering.}
For sound-hard acoustic scattering, the surface pressure $p$ satisfies the Burton--Miller equation\cite{BM71}
\begin{equation}
\bigl(\tfrac12\mathcal I-\mathcal K-\mathrm i\eta_{\mathrm a}\mathcal W\bigr)p=p^{\mathrm i}-\mathrm i\eta_{\mathrm a}\,\partial_{\hat n}p^{\mathrm i},\qquad \eta_{\mathrm a}=k,
\label{eq:bm}
\end{equation}
where $\mathcal K$ and $\mathcal W$ are the double-layer and hypersingular operators in the sign convention of Bempp-cl\cite{Bempp21}, and the imaginary coupling removes the interior resonances. The unknowns are the nodal pressures of piecewise-linear basis functions. The acoustic systems were assembled with Bempp-cl, whose time convention is $e^{-\mathrm i\omega t}$, and the operator, right-hand side and solution were complex-conjugated so that all data follow the convention of the electromagnetic problems.

\paragraph{Scaling and matrix--vector multiplications.}
All solvers are applied to the diagonally scaled system $D^{-1}Ax=D^{-1}b$ with $D=\mathrm{diag}(A)$. This scaling amounts to a diagonal preconditioner. The network is trained on this system, and the classical solvers apply their preconditioners to it. All experiments use the same discretization and differ only in how the multiplications with $A$ are computed. The dense matrix is used for problems whose matrix fits on one GPU in double precision, and an MLFMA operator\cite{SLC97} implemented on the GPU for larger electromagnetic problems. The MLFMA operator is also used for some smaller problems, which are listed in Supplementary Note~8 with the parameters of the operator and its agreement with the dense matrix. The errors of problems solved with the MLFMA operator are computed with respect to reference solutions obtained with the same operator.

\subsection*{BEMFormer}
BEMFormer consists of a network that generates three search directions in each of 25 iteration blocks, a least-squares update over the accumulated directions and a coarse-grid correction (Fig.~\ref{fig:2}a).

\paragraph{Network architecture.}
Each block consists of an input embedding, two encoder layers and an output head (Fig.~\ref{fig:2}b). The input to block $t$ is, for every basis function,
\begin{equation}
h^{(0)}=\bigl[\,r_t,\ x_t,\ b,\ Q_t\bigl(Ar_t,A\Delta x_{t-1},A\Delta x_{t-2}\bigr),\ g\,\bigr],
\label{eq:input}
\end{equation}
where complex quantities are split into real and imaginary parts and $Q_t$ projects the three operator images to eight channels. The vector $g$ contains four incident-field features, $|b|$, $\hat k\cdot\hat n$, $\cos(\mathbf k^{\mathrm i}\cdot\boldsymbol\xi)$ and $\sin(\mathbf k^{\mathrm i}\cdot\boldsymbol\xi)$, with $\boldsymbol\xi$ the position of the basis function. The 18 channels are embedded in 768 dimensions by a two-layer perceptron with layer normalization and ELU activations. The block then receives the features that the earlier blocks computed for each basis function, through a three-head attention over the block index with one learned query per block and a gated residual connection. Each encoder layer applies root-mean-square normalization before the attention described below and before a SwiGLU feed-forward layer of width 2,304\cite{Vaswani17}. The output head, a two-layer perceptron, maps the output of the last layer to the real and imaginary parts of the three search directions of the block. Each block has 17.5 million parameters, 438 million in total, and the 25 blocks do not share weights. The equations of these components are given in Supplementary Note~1.

\paragraph{Attention.}
The basis functions are partitioned into clusters of 256 by balanced k-means. Each of the eight heads of the attention combines three branches, each with its own softmax. For a basis function $n$ in cluster $\beta$ and normalized input $\tilde z=\mathrm{RMSNorm}(z)$,
\begin{equation}
\begin{gathered}
\mathrm{Attn}_n=R_{-\phi_n}\sum_{b\in\{\mathrm{loc},\mathrm{sum},\mathrm{sel}\}}\tau_{b,n}\sum_{m\in\mathcal M_b(\beta)}\operatorname*{softmax}_{m\in\mathcal M_b(\beta)}\Bigl(\frac{\hat q_n^{\top}\hat k_m}{\sqrt{128}}\Bigr)\hat v_m,\\
\tau_{b,n}=\tanh\bigl(w_b^{\top}\tilde z_n+c_b\bigr),
\end{gathered}
\label{eq:attn}
\end{equation}
where $\hat q_n=R_{\phi_n}\mathrm{RMSNorm}(q_n)$, $\hat k_m=R_{\phi_m}\mathrm{RMSNorm}(k_m)$ and $\hat v_m=R_{\phi_m}v_m$ are the rotated 128-dimensional query, key and value of the head, $q$, $k$ and $v$ are linear projections of $\tilde z$, and the head index and biases are omitted. The outputs of the eight heads are concatenated and projected back to 768 dimensions. The local branch attends to the window $\mathcal M_{\mathrm{loc}}(\beta)$ of 512 basis functions, the 256 of the cluster and the 256 nearest basis functions outside it. The summary branch attends to one key and one value for every cluster, the averages of its rotated keys and values with learned attention weights. The selection branch attends to all basis functions of two selected clusters. These are the two clusters that receive the most attention in the summary branch, summed over the queries of cluster $\beta$ and the heads, excluding $\beta$ and the clusters that supply its local window. On a quadcopter mesh with 59,256 unknowns, each cluster thus attends to 1,256 keys rather than to all 59,256 basis functions.

The operator $R_{\phi_n}$ rotates 32 of the 64 channel pairs of each head by the angles $\phi_{n,p}=k\,\mathbf d_p\cdot\boldsymbol\xi_n$ and leaves the other channels unchanged. The 32 directions $\mathbf d_p$ are distributed over the unit sphere and fixed in the frame of the incidence direction\cite{RoFormer24}. Because the rotations are orthogonal, $\hat q_n^{\top}\hat k_m$ depends on the positions only through $k\,\mathbf d_p\cdot(\boldsymbol\xi_m-\boldsymbol\xi_n)$, the plane-wave phase difference between the two points. Each value is rotated by the same phase difference.

For meshes denser than 12.5 elements per wavelength, the cluster size is set to $256\,(d/12.5)^2$ rounded to the nearest power of two, where $d$ is the mesh density in elements per wavelength. The local window is twice the cluster size. Each cluster thus covers about the same physical area as in training. For coarser meshes, the cluster size remains 256 (Supplementary Fig.~1b).

\paragraph{Least-squares update.}
The three directions of each block are normalized to unit length. The coefficients in equation~(\ref{eq:block}) are computed over all directions accumulated so far,
\begin{equation}
c_t=\arg\min_{c}\,\|r_t-A\mathcal P_tc\|^2+\mu_t\|c\|^2,
\label{eq:ls}
\end{equation}
where $\mu_t$ is $10^{-4}$ times the mean squared norm of the columns of $A\mathcal P_t$. With this regularization, the solution remains stable when the accumulated directions are nearly linearly dependent. The products $A\mathcal P_t$ are stored, so the update of the residual and the inputs $A\Delta x$ require no further matrix--vector multiplications. After 25 blocks, the solution lies in $x_0$ plus the span of 75 learned directions. One multiplication is required for the initial residual, and each of the 25 blocks requires three for $AP_t$ and one for $Ar_t$.

\paragraph{Loss function.}
The network is trained to minimize the loss
\begin{equation}
\mathcal L=\frac1B\sum_{i}\log\!\Bigl(\frac{\|b_i-Ax_{T,i}\|}{\|b_i\|}+\epsilon\Bigr)+\frac{0.1}{T}\sum_{t=1}^{T}\frac1B\sum_i\log\!\Bigl(\frac{\|b_i-Ax_{t,i}\|}{\|b_i\|}+\epsilon\Bigr),
\label{eq:loss}
\end{equation}
with $T=25$ blocks, batch size $B$ and $\epsilon=10^{-6}$. The second term penalizes the residual after every block. Gradients are backpropagated through all blocks, but the directions of a block contribute to the gradients of the least-squares solves of only the following four blocks. The loss depends only on residuals and uses no reference solution.

For an error $e=x_{\mathrm{ref}}-x$, the squared residual norm is
\begin{equation}
\|r\|^2=\|Ae\|^2=\sum_i\sigma_i^2\,|\langle v_i,e\rangle|^2,
\label{eq:svd}
\end{equation}
where $\sigma_i$ and $v_i$ are the singular values and right singular vectors of $A$. Error components along the smallest singular directions are weighted by $\sigma_i^2$ and have little effect on the loss. Training therefore reduces them only slightly. This motivates the coarse-grid correction.

\paragraph{Coarse-grid correction.}
The coarse-grid correction supplies the solution components along the near-null directions. A coarse mesh of the surface is generated with 5.5 elements per wavelength. Above about 400,000 fine unknowns, the coarse mesh density is reduced to keep about 80,000 coarse unknowns, with a minimum of 3.5 elements per wavelength. Inverse subspace iteration is started from 80 random vectors on each side and performed for 15 steps with $(A_c^{\mathrm H}A_c)^{-1}$ and $(A_cA_c^{\mathrm H})^{-1}$, both applied with the LU factors of the coarse operator $A_c$. A Rayleigh--Ritz step then yields approximations $V_{\mathrm c}$ and $U_{\mathrm c}$ of its right and left singular vectors with the smallest singular values. Only the first $k\le64$ pairs are retained, with $k$ the largest value for which the norm of the coarse-level projection $A_cV_{\mathrm c,k}(U_{\mathrm c,k}^{\mathrm H}A_cV_{\mathrm c,k})^{-1}U_{\mathrm c,k}^{\mathrm H}$ does not exceed 1.5, because a larger norm can amplify the residual. The retained vectors are interpolated to the fine mesh as $V$ and $U$.

With $\mathcal Q=AV(U^{\mathrm H}AV)^{-1}U^{\mathrm H}$, each block first computes the coefficients of the network directions on the projected residual,
\begin{equation}
y_t=\arg\min_{y}\,\|(I-\mathcal Q)(r_t-A\mathcal P_ty)\|,\qquad c_t=(U^{\mathrm H}AV)^{-1}U^{\mathrm H}(r_t-A\mathcal P_ty_t),
\label{eq:defl}
\end{equation}
and updates $x_{t+1}=x_t+\mathcal P_ty_t+Vc_t$. With the correction, the network input $x_t$ in equation~(\ref{eq:input}) is replaced by $(I-\Pi)x_t$, where $\Pi=V(U^{\mathrm H}AV)^{-1}U^{\mathrm H}A$, to remove the near-null components that the correction has added. The near-null coefficients follow from the two-sided condition $U^{\mathrm H}(r_t-A\mathcal P_ty_t-AVc_t)=0$. Testing with $U$ rather than with $AV$ makes these coefficients less sensitive to errors in the interpolated vectors. Forming $AV$ requires up to 64 matrix--vector multiplications per geometry and is shared by all incidences. The correction differs from spectral deflation\cite{CDG03} in that the vectors come from a coarse-grid operator, the cost is capped and the projection is applied inside every learned block.

For acoustic scattering, the correction follows the same procedure with the Burton--Miller operator, with the following modifications. The coarse operator is the Galerkin projection of the densely stored fine operator onto the interpolated coarse basis instead of being assembled on the coarse mesh. The number of retained vector pairs is $\min(64,\lfloor N_{\mathrm c}/2\rfloor)$, where $N_{\mathrm c}$ is the number of coarse unknowns. The interpolated vectors are corrected by one Jacobi iteration on the fine mesh (Supplementary Note~6).

\subsection*{Datasets, training and fine-tuning}
\paragraph{Electromagnetic dataset.}
The electromagnetic dataset contains 3,441 parts from the segmentation subset of the Fusion 360 Gallery dataset\cite{Lambourne21}. Each part was imported with the OpenCASCADE kernel of Gmsh, its solids were fused, and it was scaled so that its largest bounding-box dimension, drawn log-uniformly, lies between 2 and 5.5 wavelengths at 1~GHz. Each surface was meshed with target element sizes corresponding to 12 to 24 elements per wavelength, and the median mesh density of the dataset is 12.5 elements per wavelength. From the meshes with 1,000 to 30,000 unknowns, up to 238 parts were drawn from each of 15 groups. The groups combine five bins of convexity, the ratio of the enclosed volume to that of the convex hull, with three bins of the number of computer-aided design (CAD) faces. The parts of each group were split 70:10:20 into 2,408 for training, 346 for validation and 687 for testing. The training problems have 1,281--27,180 unknowns, with a 99th percentile of 19,458, and electrical sizes of 0.9--5.5 wavelengths. The electrical size is the largest dimension of the bounding box of the mesh, and it is below two wavelengths for 87 training geometries (Supplementary Note~2).

Each geometry is illuminated by vertically polarized plane waves from 72 directions ($\theta=5^\circ$--$155^\circ$ and $\varphi=0^\circ$--$330^\circ$, both in steps of $30^\circ$). The network was trained on 65 of these directions, and the remaining seven were not used for training. Reference solutions were computed by LU factorization in complex double precision and stored in single precision. Geometries were retained under two conditions. First, the condition number of $\operatorname{diag}(A)^{-1}A$ was below $10^4$. Second, at four random incidences, the combined-field solution and separate solutions of the electric- and magnetic-field integral equations had relative residuals below $10^{-5}$.

\paragraph{Training.}
The network was trained from scratch on the training set for 14 epochs with AdamW ($\beta=(0.9,0.98)$, weight decay 0.03), 384 systems per update, gradient clipping at a norm of 1 and bfloat16 mixed precision. The learning rate was linearly increased to $5\times10^{-4}$ over the first 240 updates and kept constant until the end of the tenth epoch, followed by a cosine decay to $10^{-5}$ over the last four epochs. During training, we applied a dropout rate of 0.5 after the first embedding layer and of 0.1 in the feed-forward layers. Stochastic depth with a probability of 0.15 was applied to the residual branches of the encoder layers. Training was performed on eight NVIDIA A100 and A800 GPUs with 80~GiB of memory and took 845 GPU hours, and generating the training data took 46 GPU hours (Supplementary Note~8). The training details of the models of other sizes and of the ablated models are given in Supplementary Note~3.

\paragraph{Fine-tuning for acoustic scattering.}
Fine-tuning starts from the electromagnetic weights, updates all parameters and uses the same loss, equation~(\ref{eq:loss}). The acoustic test set contains 200 geometries with 516--6,555 unknowns. From the 2,195 electromagnetic training geometries that do not share a design with these test geometries, 512 were randomly selected for fine-tuning and 64 for validation. At the acoustic frequency, 1,144~Hz, the wavelength is the same as at 1~GHz in electromagnetics, so the meshes still have 12.5 elements per wavelength. The sets with $K=8$, 32, 128 and 512 geometries are the first $K$ of the same random permutation, so that each set contains the smaller ones. Each geometry has 72 incidences, and the fine-tuning problems have up to 8,717 unknowns.

For every $K$, the network was fine-tuned for 12,288 optimizer steps, with 48 systems per update. The learning rate was linearly increased to $5\times10^{-5}$ over the first 3\% of the steps and kept constant up to half of them, followed by a cosine decay to $10^{-6}$. The checkpoint with the lowest relative error on 12 incidences of each validation geometry was used. For Fig.~\ref{fig:4}c--e, we used the network fine-tuned on the set with $K=32$, whose fine-tuning problems have up to 5,804 unknowns. For the comparison with training from scratch in Supplementary Note~6, fine-tuning and training from scratch both used 768 optimizer steps, with the same architecture and data. The learning-rate schedule was the same, with its phases shortened in proportion to the number of steps, and the peak learning rate was $5\times10^{-4}$ for training from scratch.

\paragraph{Fine-tuning for the electric-field integral equation.}
The electric-field integral equation is equation~(\ref{eq:cfie}) with $\alpha=1$. The fine-tuning dataset contains 297 geometries of the electromagnetic training set with 72 incidences each. The network was fine-tuned in the same way for three epochs with 126 systems per update. The learning rate was linearly increased to $2\times10^{-5}$ over the first 2\% of the steps, followed by a cosine decay to $2\times10^{-6}$. The fine-tuned network was tested on the corresponding out-of-distribution set of 199 geometries with eight incidences each. The coarse-grid correction is not used for this equation (Supplementary Note~5).

\subsection*{Evaluation}
\paragraph{Metrics.}
The relative residual is $\|b-Ax\|/\|b\|$ and the relative error is $\|x-x_{\mathrm{ref}}\|/\|x_{\mathrm{ref}}\|$. The reference $x_{\mathrm{ref}}$ is the dense LU solution. For problems too large for dense LU, it is the solution of unrestarted block-Jacobi GMRES with the same MLFMA operator. This solution has a relative residual below $3.2\times10^{-6}$, or of about $10^{-5}$ where the Krylov basis had to be stored in single precision. The RCS is $\sigma=\lim_{R\to\infty}4\pi R^2|\mathbf E^{\mathrm s}|^2/|\mathbf E^{\mathrm i}|^2$, and the monostatic RCS is observed in the direction from which the wave is incident. The RCS is expressed in decibels relative to one square metre (dBsm). The mean RCS deviation over a set $\Theta$ of observation angles is
\begin{equation}
\overline D=\frac1{|\Theta|}\sum_{\varphi\in\Theta}\bigl|\sigma_{\mathrm{dB}}(\varphi)-\sigma_{\mathrm{ref,dB}}(\varphi)\bigr|,
\label{eq:rcs}
\end{equation}
taken over the full curve or pattern, without excluding any angle. The 1~dB threshold is based on the two independent measurements of the almond. Under this metric, they differ by a median of 1.01~dB over eight combinations of frequency and polarization. Measurement facilities quote similar uncertainties\cite{AustinAPS22}, although they state them as pointwise intervals. The relative error of $x$ also measures the error of the surface current (Supplementary Note~4). The near-null error fraction is $f_8=\sum_{i>N-8}|\langle v_i,e\rangle|^2/\|e\|^2$. In Figs.~\ref{fig:2}e, \ref{fig:3}e and \ref{fig:6}a,g, surface currents and their errors are shown in dB, normalized to the 99th percentile of the magnitude of the reference current.

\paragraph{Out-of-distribution test sets.}
The out-of-distribution sets are derived from 200 test parts, drawn from each group in proportion to its share of the training set. Six sets use the test meshes. They are a reference set with the training settings and sets with horizontal polarization, combination coefficients $\alpha=0.7$ and 0.3, the magnetic-field integral equation alone ($\alpha=0$) and the electric-field integral equation ($\alpha=1$). For five other sets, the parts were meshed again at the same physical size, for 0.5 and 2~GHz and for 1~GHz with about 10, 16.5 and 24.5 elements per wavelength. Each set contains 110 to 200 geometries and is evaluated at eight incidences, seven of which are not among the training incidences. Two additional sets replace the plane wave with an electric dipole or an infinite line current. They use the geometries of the reference set, and its eight incidences serve as nominal incidence directions. The inputs of the network that depend on the incidence direction are computed for the nominal plane wave, so only the right-hand side changes. The construction of all sets is described in Supplementary Note~2, and the results for each out-of-distribution set are given in Supplementary Note~5.

\paragraph{Baselines.}
All classical solvers start from $x_0=b$. Dense LU is performed with cuSOLVER in complex double precision. GMRES is used without restarting, and each new Krylov vector is orthogonalized against all previous ones\cite{SaadSchultz86}. The block-Jacobi preconditioner inverts the diagonal blocks of a k-means partition of the basis functions into clusters of about 256 unknowns. The near-field preconditioner is the sparse LU factorization of the near-field part of the operator\cite{MalasGurel07}. The near-field part consists of the entries between basis functions in the same or adjacent cubes of 0.4 wavelengths, the finest boxes of the MLFMA. For the largest quadcopter meshes, the near-field matrix was truncated to fit in memory (Supplementary Note~8). The biconjugate gradient stabilized method (BiCGSTAB) is included as a short-recurrence alternative to GMRES (Supplementary Note~4). Extended Data Fig.~\ref{ed:3} compares BEMFormer with GMRES and block-Jacobi GMRES deflated by the same coarse space, and with near-field-LU GMRES.

The learned baselines were trained on the same training set and evaluated with the same metrics, following ref.~\cite{McGreivy24}. Their evaluation uses eight of the 72 incidences of each test geometry, 5,496 systems in total. Transolver\cite{Transolver24} and PhiGRL\cite{PhiGRL24} predict the solution in one forward pass. Transolver was trained in two sizes, the original model with 3.9 million parameters and an enlarged model with 439 million parameters. We reimplemented the PdEgatSCL model\cite{PdEgatSCL25} from the code released by its authors and trained it with their loss function and optimizer settings.

\paragraph{Measurements and analytical reference.}
The almond measurements at 3.5, 5.125, 7 and 10.25~GHz are the two data sets of ref.~\cite{Kelley20}, distributed with the Austin RCS benchmark suite\cite{Massey18}. The ARCHIE-AIM results are the reference solutions distributed with the suite. The measurements of the almond at 1.19 and 9.92~GHz and of the ogive, double ogive and cone-sphere are those of refs.~\cite{Woo93,Gibson08}. These curves were extracted from published figures, and the uncertainty of their angular alignment is about $1^\circ$ (Supplementary Note~7). For the sphere, the reference is the Mie series\cite{BohrenHuffman83}. Discretization error is estimated by comparing meshes with 12.5 and 25 elements per wavelength. For the almond at 3.5~GHz, meshes with 37.5 and 50 elements per wavelength were added and solved by dense LU. The combined-field integral equation is known to be less accurate than the electric-field integral equation with Rao--Wilton--Glisson functions\cite{ErgulGurel06,ErgulGurel09}.

\paragraph{Timing and hardware.}
End-to-end time comprises matrix assembly, the one-time preparation of each method, the solutions and post-processing. For the MLFMA operator, matrix assembly includes the octree, the translation operators, the radiation and receiving patterns of the basis functions and the near-field matrix. For $N_{\mathrm{inc}}$ incidences, the time is $t=t_{\mathrm{prep}}+N_{\mathrm{inc}}\,t_{\mathrm{solve}}$, so the speed-up approaches the ratio of solution times as $N_{\mathrm{inc}}$ increases. Times for $N_{\mathrm{inc}}>1$ were computed from the measured one-time cost and the median measured time per incidence. Each classical solver was run, for every incidence, until it reached the RCS deviation that BEMFormer attained for that incidence. All methods were timed on the same GPU in three interleaved repetitions, with no other job on the machine, and the median is reported. The GPUs were NVIDIA A100 80GB PCIe at their default power limit of 300~W, in a server with two Intel Xeon Gold 6330 CPUs.

\paragraph{Statistics.}
We report medians, the number of systems or geometries on which one method is better than another, and 95\% confidence intervals for fitted slopes. The slopes of solution time against the number of unknowns were obtained by linear regression of the logarithms, and their 95\% confidence intervals were estimated using Student's $t$-distribution (Supplementary Note~8).

\subsection*{Use of large language models}
Large language models (Claude, Anthropic; Codex, OpenAI) were used to assist with coding and with language editing and refinement of the Article text. All code was reviewed and tested by the authors. All scientific content, interpretations and conclusions were developed by the authors.

\backmatter

\bmhead{Supplementary information}
The Supplementary Information contains Supplementary Notes~1--9, Supplementary Figs.~1--6 and Supplementary Tables~1--8.

\begin{figure}[p]
\centering
\thisfloatpagestyle{empty}\makebox[\textwidth][c]{\includegraphics[width=180mm]{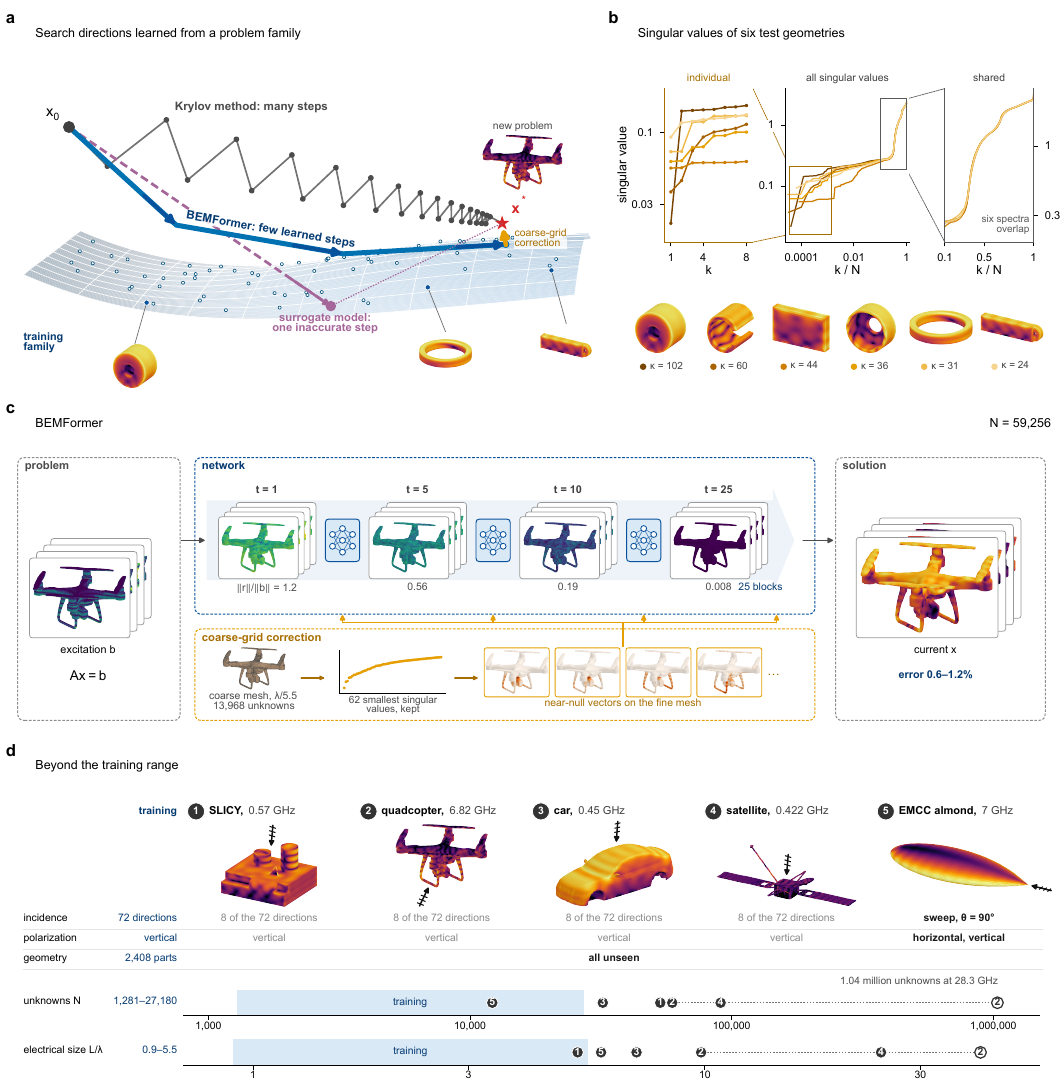}}
\caption{\textbf{BEMFormer learns to generate search directions for a family of scattering problems.} \textbf{a}, Schematic of the solution space of a family of scattering problems, with the paths of a Krylov method, a surrogate model and BEMFormer. \textbf{b}, Singular values of six test geometries versus the index $k$ divided by the number of unknowns $N$ (middle). Magnified views show the eight smallest values (left) and the values with $k/N\ge0.1$ (right). The six geometries are shown below with their condition numbers $\kappa$. \textbf{c}, Solution procedure for the quadcopter, with the residual at the start of four blocks. \textbf{d}, Test targets outside the training distribution. Numbered circles indicate the number of unknowns $N$ and the electrical size $L/\lambda$ of each target, and shaded bars indicate the training range. Colours indicate the magnitude of the surface current. Open circles indicate the quadcopter at 28.3~GHz. Arrows indicate the incidence direction, and the short bars across them are parallel to the incident electric field.}\label{fig:1}
\end{figure}

\begin{figure}[p]
\centering
\thisfloatpagestyle{empty}\makebox[\textwidth][c]{\includegraphics[width=180mm]{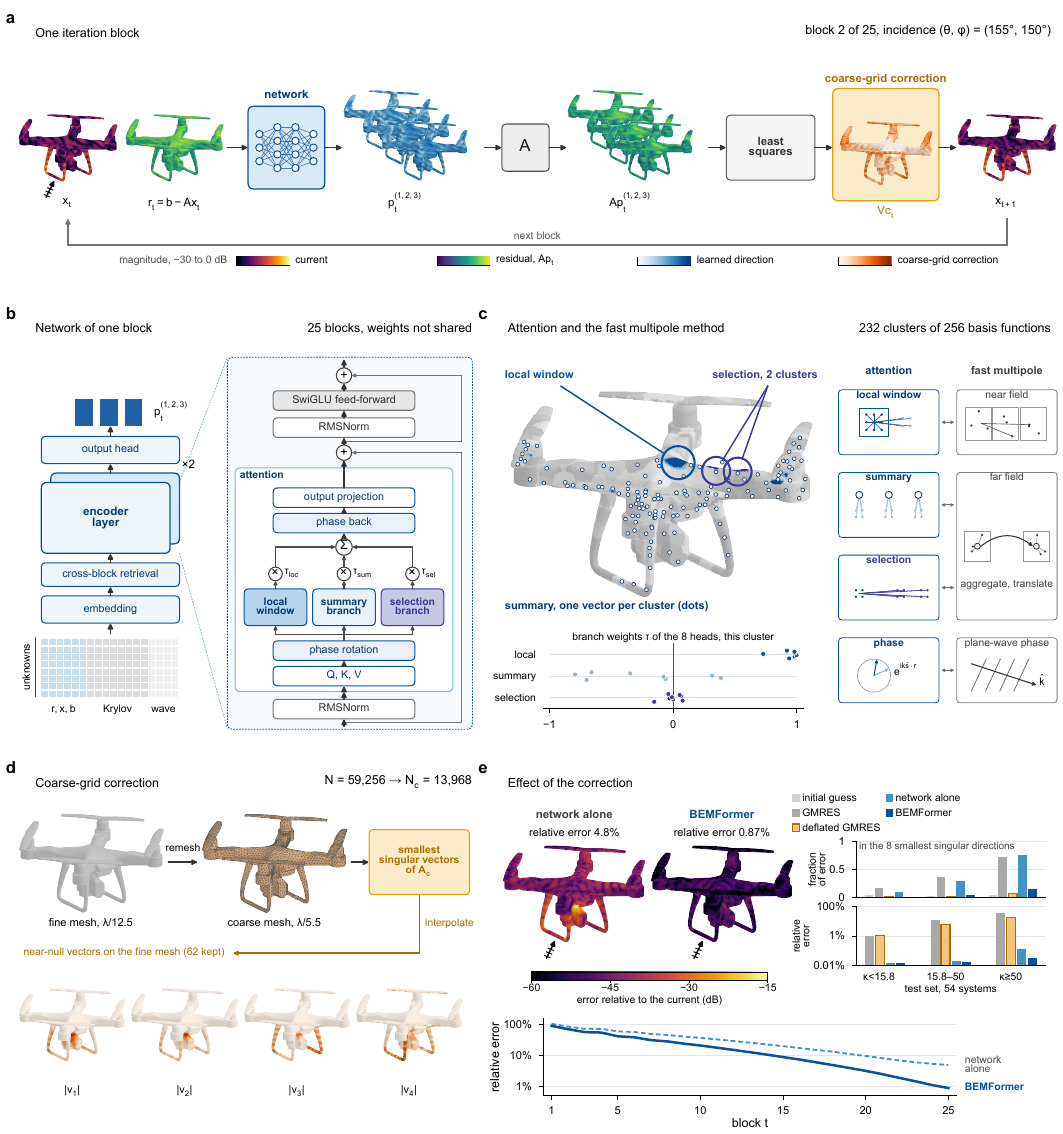}}
\caption{\textbf{The BEMFormer iteration.} \textbf{a}, One iteration block for the quadcopter. The maps show the solution $x_t$, the residual $r_t$, the three directions $p_t$ generated by the network, their products with the operator $Ap_t$ and the coarse-grid correction $Vc_t$. \textbf{b}, Network of one block (left) and one of its encoder layers (right). \textbf{c}, Attention branches for one cluster of the quadcopter mesh (left) and their counterparts in the fast multipole method (right). \textbf{d}, Coarse-grid correction. Near-null vectors of the operator on a coarse mesh are interpolated to the fine mesh, and the first four are shown. \textbf{e}, Effect of the correction. Left, error of the network alone and of BEMFormer for the incidence of \textbf{a}. Right, median fraction of the error in the eight smallest singular directions (top) and median relative error (bottom) for 54 test systems. GMRES and deflated GMRES were run for 101 matrix--vector multiplications. Bottom, relative error after each block. Arrows indicate the incidence direction and polarization, as in Fig.~\ref{fig:1}d.}\label{fig:2}
\end{figure}

\begin{figure}[p]
\centering
\thisfloatpagestyle{empty}\makebox[\textwidth][c]{\includegraphics[width=180mm]{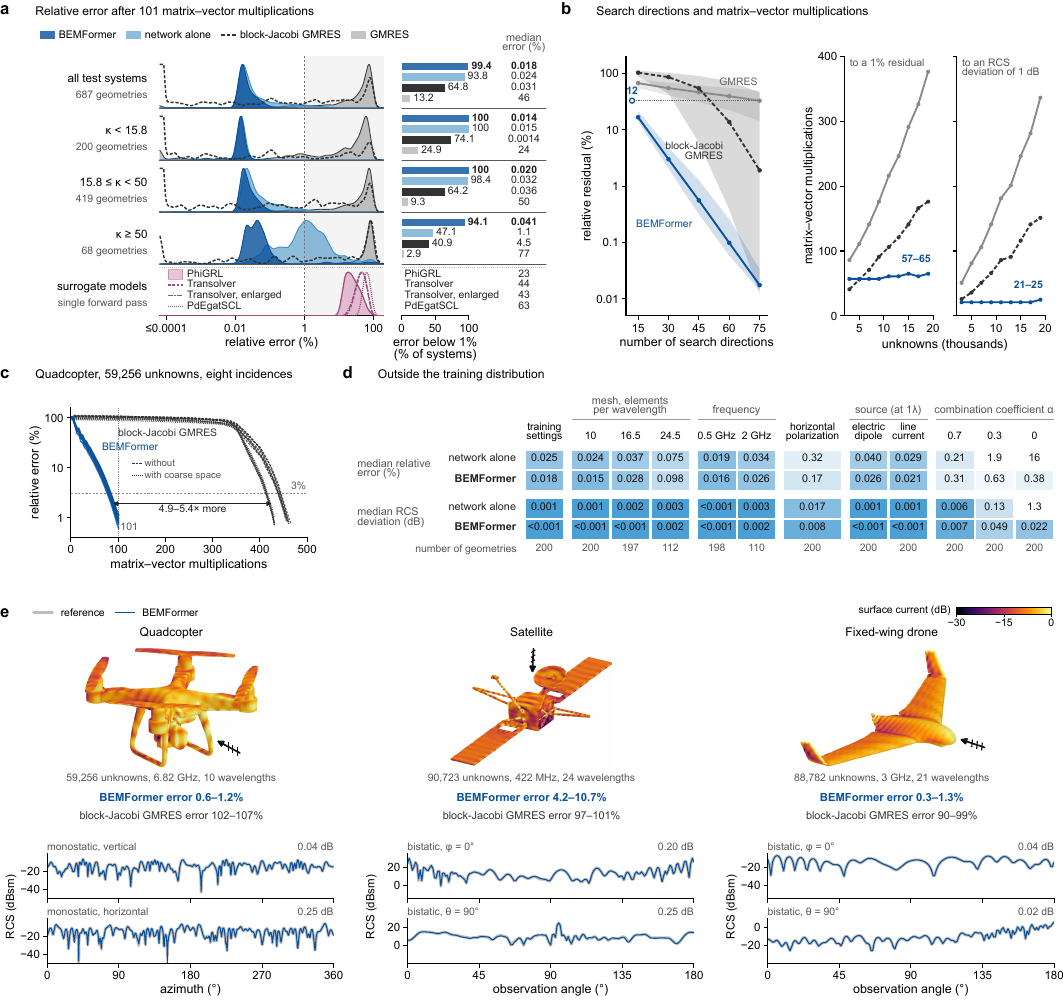}}
\caption{\textbf{Accuracy and number of search directions on held-out electromagnetic problems.} \textbf{a}, Distribution of the relative error after 101 matrix--vector multiplications on the test set ($n=49{,}464$ systems), for all systems and grouped by the condition number $\kappa$. The bottom row shows the surrogate models after a single forward pass ($n=5{,}496$), with Transolver in its original size and enlarged to 439 million parameters. Bars indicate the percentage of systems with an error below 1\%, and the numbers on the right indicate the median relative error. GMRES uses a diagonal preconditioner. \textbf{b}, Left, relative residual versus the number of search directions for 2,061 test systems, with medians (lines) and interquartile ranges (bands). The open circle indicates the point at which BEMFormer reaches the residual of GMRES after 75 iterations. Right, median number of matrix--vector multiplications required to reach a 1\% residual and an RCS deviation of 1~dB, in bins of 2,000 unknowns. \textbf{c}, Relative error versus the number of matrix--vector multiplications on the quadcopter. \textbf{d}, Median relative error and median RCS deviation on 11 test sets outside the training distribution. The first column shows the training settings. In the source columns, an electric dipole or a line current at a distance of one wavelength from the target replaces the plane wave. \textbf{e}, Three targets outside the training distribution. Top, surface current of the BEMFormer solution (arrows as in Fig.~\ref{fig:1}d), with the range of the relative error over eight incidences. Bottom, RCS with the mean deviation from the reference.}\label{fig:3}
\end{figure}

\begin{figure}[p]
\centering
\makebox[\textwidth][c]{\includegraphics[width=180mm]{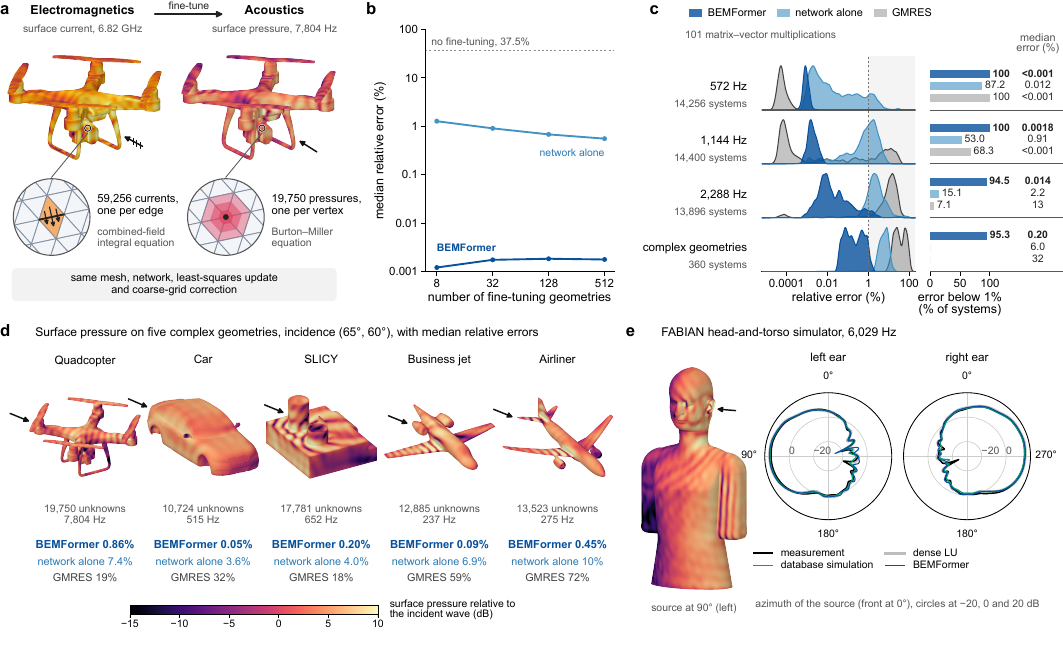}}
\caption{\textbf{Few-shot transfer to acoustic scattering.} \textbf{a}, Unknowns of the two problems on the same mesh, one current per edge in electromagnetics and one pressure per vertex in acoustics. \textbf{b}, Median relative error on the acoustic test set versus the number of fine-tuning geometries. \textbf{c}, Distribution of the relative error after 101 matrix--vector multiplications at three frequencies and on five complex geometries. Bars indicate the percentage of systems with an error below 1\%, and the numbers on the right indicate the median relative error. \textbf{d}, Surface pressure of the BEMFormer solution on the five complex geometries. Numbers indicate the median relative error of each method over 72 incidences. \textbf{e}, FABIAN head-and-torso simulator. Left, surface pressure of the BEMFormer solution. Right, head-related transfer functions of both ears versus the azimuth of the source. The colour scale is the same as in \textbf{d}. The network in \textbf{c}--\textbf{e} was fine-tuned on 32 geometries. Arrows indicate the incidence direction, with short bars as in Fig.~\ref{fig:1}d.}\label{fig:4}
\end{figure}

\begin{figure}[p]
\centering
\makebox[\textwidth][c]{\includegraphics[width=180mm]{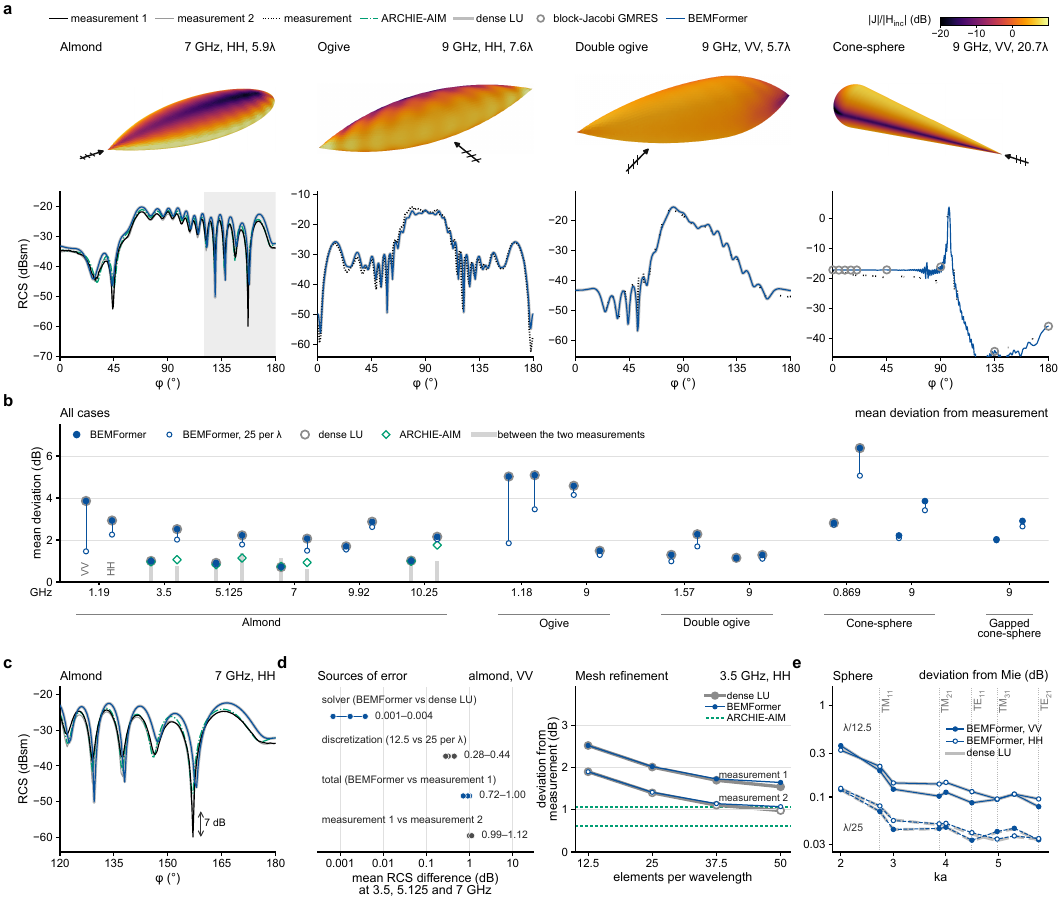}}
\caption{\textbf{Validation against measurements and an analytical solution.} \textbf{a}, Surface current of the BEMFormer solution (top) and monostatic RCS versus the azimuth $\varphi$ (bottom) for four EMCC targets. Arrows indicate the incidence direction and polarization, as in Fig.~\ref{fig:1}d. \textbf{b}, Mean RCS deviation from measurement for all cases. \textbf{c}, Magnified view of the shaded region in \textbf{a}. The arrow indicates the difference between the two measurements at a null. \textbf{d}, Sources of error for the almond (left) and deviation from the two measurements versus the mesh density (right). \textbf{e}, Mean RCS deviation from the Mie series for a perfectly conducting sphere of radius $a$ at wavenumber $k$. Vertical lines indicate the interior resonances. Measurements 1 and 2 are the two independent measurements of the almond, and the other measurements were obtained from refs.~\cite{Woo93,Gibson08}. VV and HH denote vertical and horizontal polarization.}\label{fig:5}
\end{figure}

\begin{figure}[p]
\centering
\thisfloatpagestyle{empty}\makebox[\textwidth][c]{\includegraphics[width=180mm]{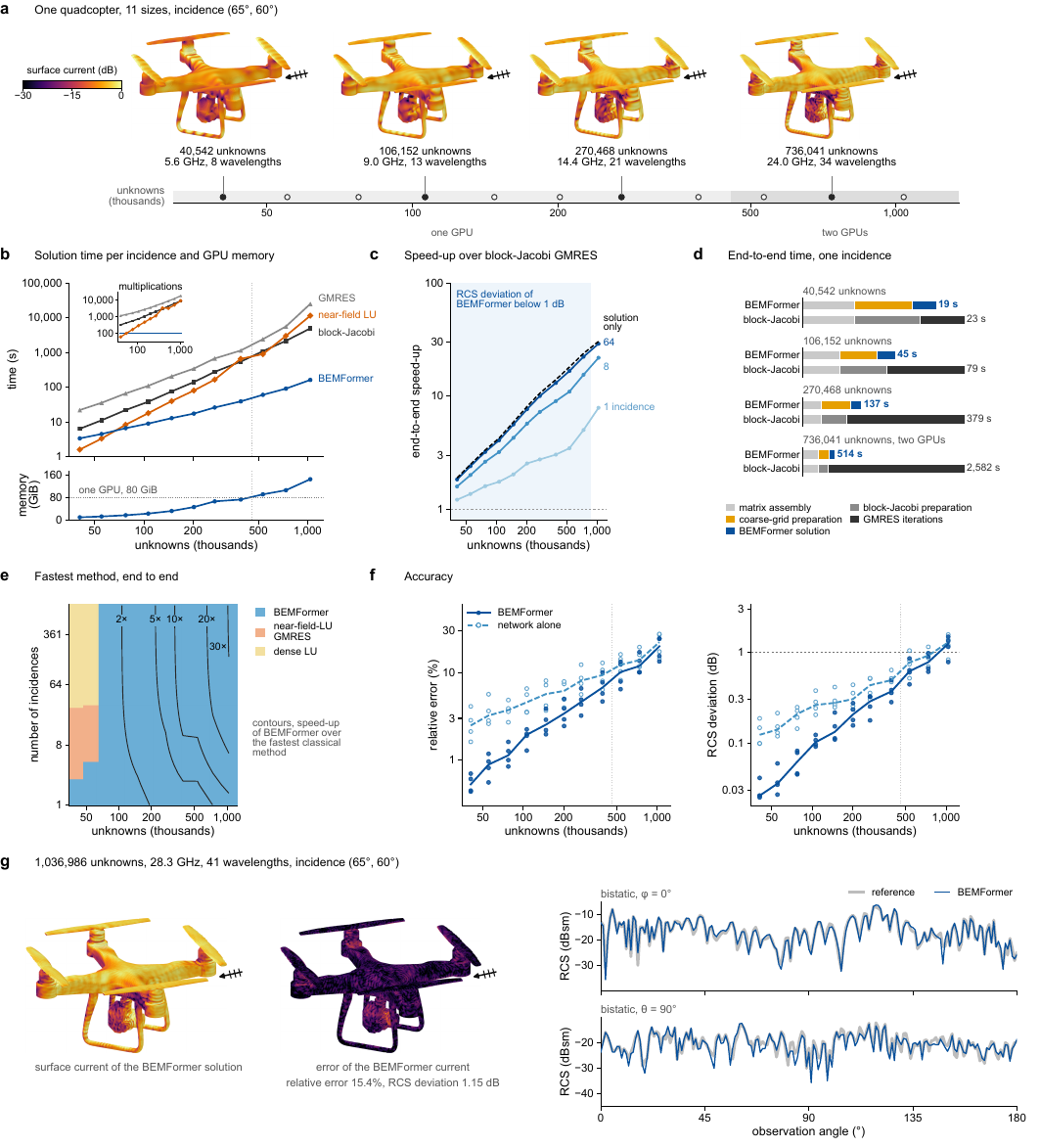}}
\caption{\textbf{Cost and accuracy on a quadcopter with up to 1.04 million unknowns.} \textbf{a}, Surface current of the BEMFormer solution on four of the 11 meshes. Arrows indicate the incidence direction and polarization, as in Fig.~\ref{fig:1}d. \textbf{b}, Median solution time per incidence (top) and graphics processing unit (GPU) memory of BEMFormer (bottom) versus the number of unknowns. The inset shows the number of matrix--vector multiplications that each method requires to reach the RCS deviation of BEMFormer. \textbf{c}, End-to-end speed-up over block-Jacobi GMRES. \textbf{d}, End-to-end time for one incidence at the four sizes of \textbf{a}. \textbf{e}, Fastest method for each number of unknowns and of incidences. \textbf{f}, Relative error (left) and RCS deviation (right) versus the number of unknowns. Points show the four incidences at each size and lines show their medians. \textbf{g}, Surface current of the BEMFormer solution (left), its error (middle) and bistatic RCS (right) at 1.04 million unknowns. The colour scale is the same as in \textbf{a}. Vertical dotted lines in \textbf{b} and \textbf{f} separate the sizes solved on one and on two GPUs.}\label{fig:6}
\end{figure}

\bmhead{Data availability}
The meshes of the training, test, out-of-distribution and acoustic test sets have been deposited in Zenodo (ref.~\cite{SunData26}), together with the reference solutions of the test set and the trained models. They were derived from the Fusion 360 Gallery dataset\cite{Lambourne21} and may be used for non-commercial research only. The meshes and solutions of the quadcopter series, the fixed-wing drone, the satellite, the high-speed train, the EMCC targets and the sphere have been deposited in Zenodo (ref.~\cite{SunCode26}). Both records will be made publicly available upon publication. The geometry of the car in Figs.~\ref{fig:1}d and \ref{fig:4}d was derived from the DrivAerML dataset\cite{Ashton24}. The geometries of the satellite, the airliner and the business jet were derived from NASA models.

\bmhead{Code availability}
The code of BEMFormer is available via GitHub at \url{https://github.com/shrrr/BEMFormer}. The repository also contains the source data for all figures. The version used in this study has been deposited in Zenodo (ref.~\cite{SunCode26}) and will be made publicly available upon publication.

\bmhead{Acknowledgements}
This work was supported by the National Key Research and Development Program of China (grant no. 2025YFC2423102). Computing resources were provided by the High Performance Computing Center, Tsinghua University.

\bmhead{Author contributions}
H.S., T.S. and M.L. conceived the study. H.S. developed the method, implemented the code, performed the experiments and wrote the manuscript. T.S. helped with the development of the MLFMA operator. H.Z. and R.G. contributed to discussions regarding the method and the results. S.X., F.Y. and M.L. supervised the project and acquired funding. All authors discussed the results and commented on the manuscript.

\bmhead{Competing interests}
The authors declare no competing interests.

\bibliography{main}

\clearpage

\clearpage
\setcounter{figure}{0}
\renewcommand{\figurename}{Extended Data Fig.}

\begin{figure}[p]\centering
\makebox[\textwidth][c]{\includegraphics[width=180mm]{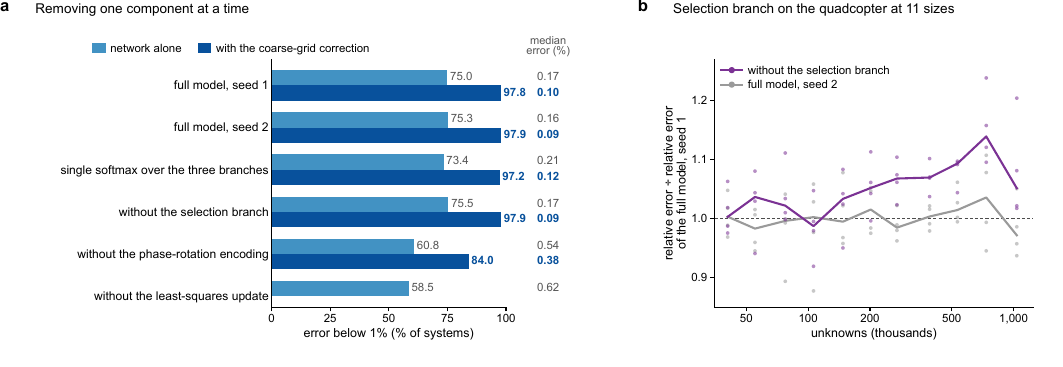}}
\caption{\textbf{Ablation of the network components.} \textbf{a}, Percentage of test systems with a relative error below 1\% ($n=5{,}496$ systems) for the full model trained with two random seeds and for four ablated models. The numbers on the right indicate the median relative error. The model without the least-squares update is shown for the network alone. \textbf{b}, Relative error on the quadcopter at 11 sizes, divided by the relative error of the full model with the first seed. The curves show the model without the selection branch and the full model with the second seed. Points show the four incidences at each size and lines show their medians. All models have 19.3 million parameters and were trained for eight epochs.}\label{ed:2}\end{figure}

\begin{figure}[p]\centering
\makebox[\textwidth][c]{\includegraphics[width=180mm]{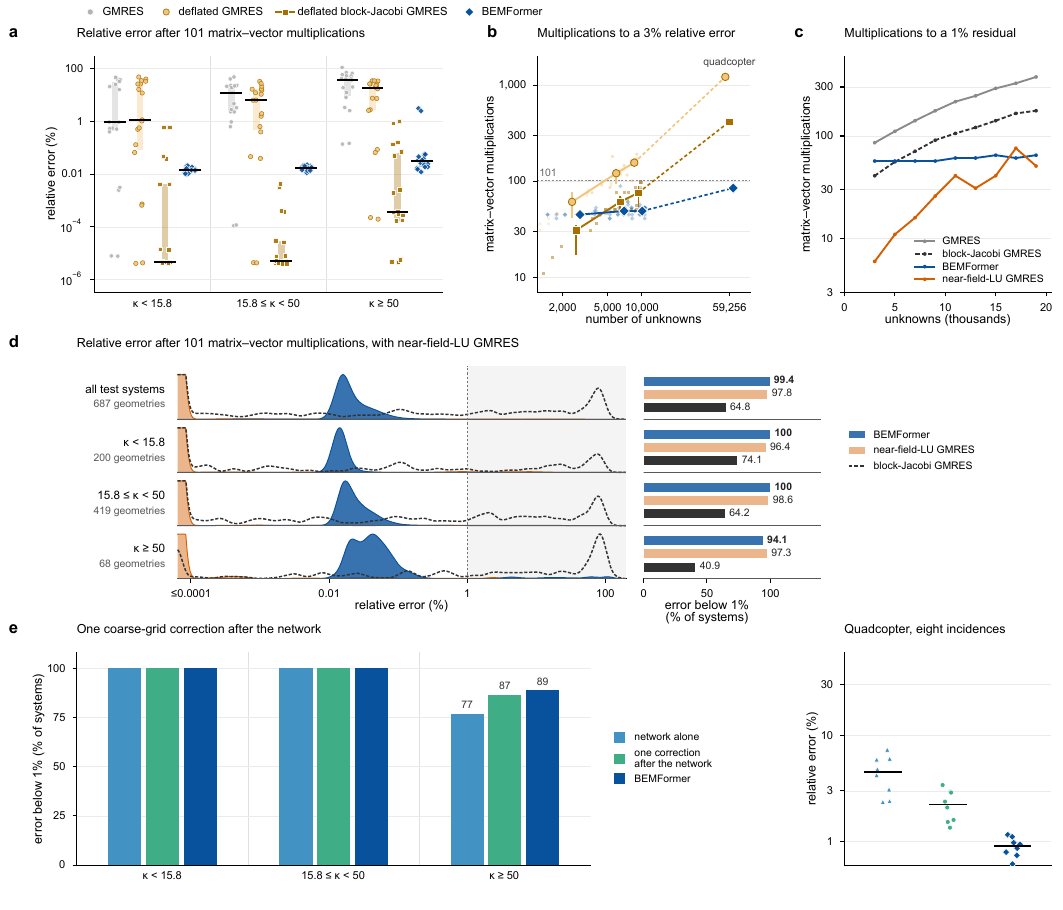}}
\caption{\textbf{Classical Krylov solvers with the coarse space of BEMFormer or a near-field preconditioner.} \textbf{a}, Relative error after 101 matrix--vector multiplications for 54 test systems grouped by the condition number $\kappa$. For deflated GMRES and deflated block-Jacobi GMRES, the deflation space is the coarse space of BEMFormer. Shaded bars indicate the interquartile range and black lines indicate the median. \textbf{b}, Number of matrix--vector multiplications required to reach a 3\% relative error versus the number of unknowns $N$. Large markers indicate the medians in three ranges of $N$ and on the quadcopter, vertical bars indicate the interquartile range and small markers indicate individual systems. \textbf{c}, Median number of matrix--vector multiplications required to reach a 1\% residual on the test set, in bins of 2,000 unknowns. \textbf{d}, Distribution of the relative error after 101 matrix--vector multiplications on the test set, as in Fig.~\ref{fig:3}a, with near-field-LU GMRES. Bars indicate the percentage of systems with an error below 1\%. \textbf{e}, One coarse-grid correction applied after the network instead of in every block. Left, percentage of test systems with a relative error below 1\%. Right, relative error for eight incidences on the quadcopter. Black lines indicate the median.}\label{ed:3}\end{figure}

\begin{figure}[p]\centering
\thisfloatpagestyle{empty}\makebox[\textwidth][c]{\includegraphics[width=180mm]{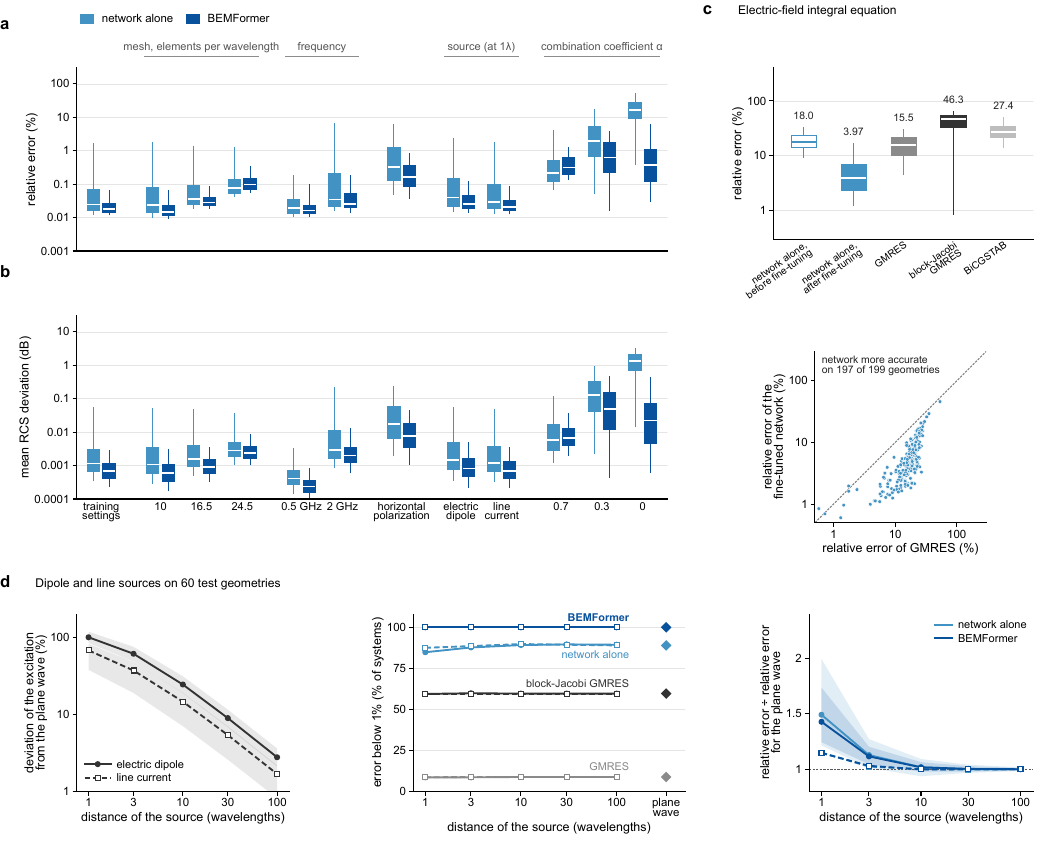}}
\caption{\textbf{Out-of-distribution tests.} \textbf{a},\textbf{b}, Relative error (\textbf{a}) and mean RCS deviation (\textbf{b}) after 101 matrix--vector multiplications for the training settings and for 11 test sets outside the training distribution. Boxes indicate the interquartile range, centre lines indicate the median and whiskers indicate the 5th and 95th percentiles. Each set has 110--200 geometries and eight incidences. \textbf{c}, Electric-field integral equation ($n=1{,}592$ systems of 199 geometries). Top, relative error after 101 matrix--vector multiplications for the network alone before and after fine-tuning and for three classical solvers. Numbers indicate the median. Bottom, mean relative error of the fine-tuned network versus the mean relative error of GMRES. Each point represents one geometry. \textbf{d}, Electric dipole and line current at 1 to 100 wavelengths from the target, on 60 test geometries ($n=480$ systems for each source and distance). Left, deviation of the excitation from the plane wave. Middle, percentage of systems with a relative error below 1\%. Right, relative error divided by the relative error of the same system for the plane wave. Lines show medians and bands show interquartile ranges, on the right for the dipole only.}\label{ed:5}\end{figure}
\end{document}